\documentclass[%
 reprint,
superscriptaddress,
 amsmath,amssymb,
 aps,
floatfix,
]{revtex4-2}

\usepackage{graphicx}
\usepackage{dcolumn}
\usepackage{bm}
\usepackage{amsmath,amssymb}
\usepackage{textcomp,mathcomp}
\usepackage{color}
\begin{document}

\preprint{APS/123-QED}

\title{Reexamination of the charge-ordered dimer pattern in the spinel compound CuIr$_2$S$_4$ using single-crystal synchrotron x-ray diffraction}

\author{T. Ohashi}
\affiliation{Department of Applied Physics, Nagoya University, Nagoya 464-8603, Japan}
\author{N. Katayama}\thanks{Corresponding author.}\email{katayama.naoyuki.m5@f.mail.nagoya-u.ac.jp}
\affiliation{Department of Applied Physics, Nagoya University, Nagoya 464-8603, Japan}
\author{K. Kojima}
\affiliation{The Insitute for Solid State Physics, The University of Tokyo, Kashiwa 277-8581, Japan}
\author{M. Emi}
\affiliation{Department of Applied Physics, Nagoya University, Nagoya 464-8603, Japan}
\author{C. Koyama}
\affiliation{Department of Applied Physics, Nagoya University, Nagoya 464-8603, Japan}
\author{T. Hara}
\affiliation{Department of Applied Physics, Nagoya University, Nagoya 464-8603, Japan}
\author{K. Hashimoto}
\affiliation{Materials and Structures Laboratory, Institute of Science Tokyo, 4259 Nagatsuta-cho, Midori-ku, Yokohama 226-8503, Japan}
\author{S. Kitani}
\affiliation{Materials and Structures Laboratory, Institute of Science Tokyo, 4259 Nagatsuta-cho, Midori-ku, Yokohama 226-8503, Japan}
\author{H. Kawaji}
\affiliation{Materials and Structures Laboratory, Institute of Science Tokyo, 4259 Nagatsuta-cho, Midori-ku, Yokohama 226-8503, Japan}
\author{H. S. Suzuki}
\affiliation{The Insitute for Solid State Physics, The University of Tokyo, Kashiwa 277-8581, Japan}
\author{S. Nagata}
\affiliation{Department of Materials Science and Engineering, Muroran Institute of Technology, Mizumoto-cho, Muroran, Hokkaido 050-8585, Japan}
\author{K. Sugimoto}
\affiliation{Department of Physics, Keio University, Kohoku-ku, Yokohama, Kanagawa 223-8522, Japan}
\author{K. Iida}
\affiliation{Neutron Science and Technology Center, Comprehensive Research Organization for Science and Society (CROSS), Tokai, Ibaraki 319-1106, Japan}
\author{H. Sawa}					
\affiliation{Department of Applied Physics, Nagoya University, Nagoya 464-8603, Japan}
\date{\today}

\begin{abstract}
We have re-investigated the crystal structure of a spinel type CuIr$_2$S$_4$ at low temperatures using a single-crystal in a synchrotron radiation x-ray diffraction experiment. The crystal structure of the low-temperature phase of CuIr$_2$S$_4$ has been already studied by diffraction experiments using a powder sample, and it has been reported that the formation of dimer molecules accompanied by charge ordering of Ir has been achieved. The crystal structure of the low-temperature phase obtained in our reanalysis was the same as the previously reported structure in that it showed the formation of Ir dimers accompanied by charge ordering, but the charge ordering pattern and arrangement of the dimers in the unit cell were different. We will discuss the validity of the structure obtained in this study and provide the structural parameters revealed in the reanalysis. The results of this study should provide a basis for further studies of the physical properties of CuIr$_2$S$_4$, which are still being actively investigated.
\end{abstract}

\maketitle


The exploration of novel ground states in which the multiple electronic degrees of freedom coupled with the lattice is a central topic in condensed matter physics. An instance of such a ground state is the molecular formation phenomenon that occurs when transition metal ions spontaneously aggregate \cite{CsW2O6,MgTi2O4,AlV2O4,Fe3O4,RuP, Li033VS2,M2,VO2_Kitou, BCSO, BCSO-2, GaNb4Se8, LiMoO2,In2Ru2O7, LiVX2_review, LiVO2-kj2, LiVS2-1, LiVS2-2, LiRh2O4, LiRh2O4-2, CuIr2S4_radaelli,CuIr2S4_xrayinduced, CuIr2S4_xrayinduced2,CuIr2S4_rixs,CuIr2S4_incomme}. Examples include the vanadium trimerization that occurs in the layered triangular lattice systems LiVO$_2$ and LiVS$_2$ \cite{LiVO2_and_LiVS2_trimerization, LiVX2_review, LiVO2-kj2, LiVS2-1, LiVS2-2}, and the dimer formation that accompanies the charge ordering state that occurs in the spinel lattice systems LiRh$_2$O$_4$ \cite{LiRh2O4, LiRh2O4-2} and CuIr$_2$S$_4$ \cite{CuIr2S4_MItransition,CuIr2S4_xrayinduced, CuIr2S4_xrayinduced2,CuIr2S4_rixs,CuIr2S4_incomme,MizoKom}. In particular, CuIr$_2$S$_4$ continues to provide fascinating topics even today, such as the melting phenomenon of the charge-ordered/dimer state that appears under high-intensity x-ray irradiation \cite{CuIr2S4_xrayinduced, CuIr2S4_xrayinduced2,CuIr2S4_rixs,CuIr2S4_incomme,manganite_xrayinduced,VO2_xrayinduced}, novel magnetic properties that appear below 100 K due to spin-orbit interaction \cite{CuIr2S4_muonspin,CuIr2S4_nasu}, local tetragonal distortion that appears prior to molecular formation in the high-temperature lattice \cite{CuIr2S4_HT_shortrangeorder}, and high-temperature superconductivity with the highest $T_c$ = 18.2 K that appears by suppressing the molecular formation under high pressure \cite{CuIr2S4_superconductivity}. The background to the appearance of the diverse phenomena described above in CuIr$_2$S$_4$ includes the charge degree of freedom of the half-integer Ir, the geometric frustration effect caused by the pyrochlore lattice, the orbital degeneracy on the pyrochlore lattice with high symmetry, and the strong spin-orbit interaction derived from the 5$d$ element. By precisely examining the structure of the low-temperature phase that is formed when these factors interact and crystallize, we should be able to obtain important clues for understanding these phenomena.

For the low-temperature phase of CuIr$_2$S$_4$, it was reported in 2002 that an Ir$^{3+}$/Ir$^{4+}$ charge-ordered state occurs on the pyrochlore lattice of Ir, and dimer formation occurs between neighboring Ir$^{4+}$ ions \cite{CuIr2S4_radaelli}. In this literature, the crystal structure of the low-temperature phase was investigated using a combination of electron diffraction, neutron diffraction, and x-ray diffraction techniques, and the structure of the low-temperature phase was identified by Rietveld analysis of the diffraction data for the powder sample. The low-temperature phase of CuIr$_2$S$_4$ belongs to the triclinic $P\bar{1}$ space group, and all 28 crystallographically independent sites in the unit cell are in general positions. In general, precise Rietveld analysis of powder diffraction data for such low-symmetry structures is extremely difficult. As mentioned above, CuIr$_2$S$_4$ is still an attractive research subject, so re-analysis of the low-temperature phase structure using single-crystal samples is still an important issue that should be addressed. The precise crystal structure data obtained through re-analysis using a single-crystal sample should provide a new basis for studying many topics related to CuIr$_2$S$_4$.

\begin{figure*}
\includegraphics[width=180mm]{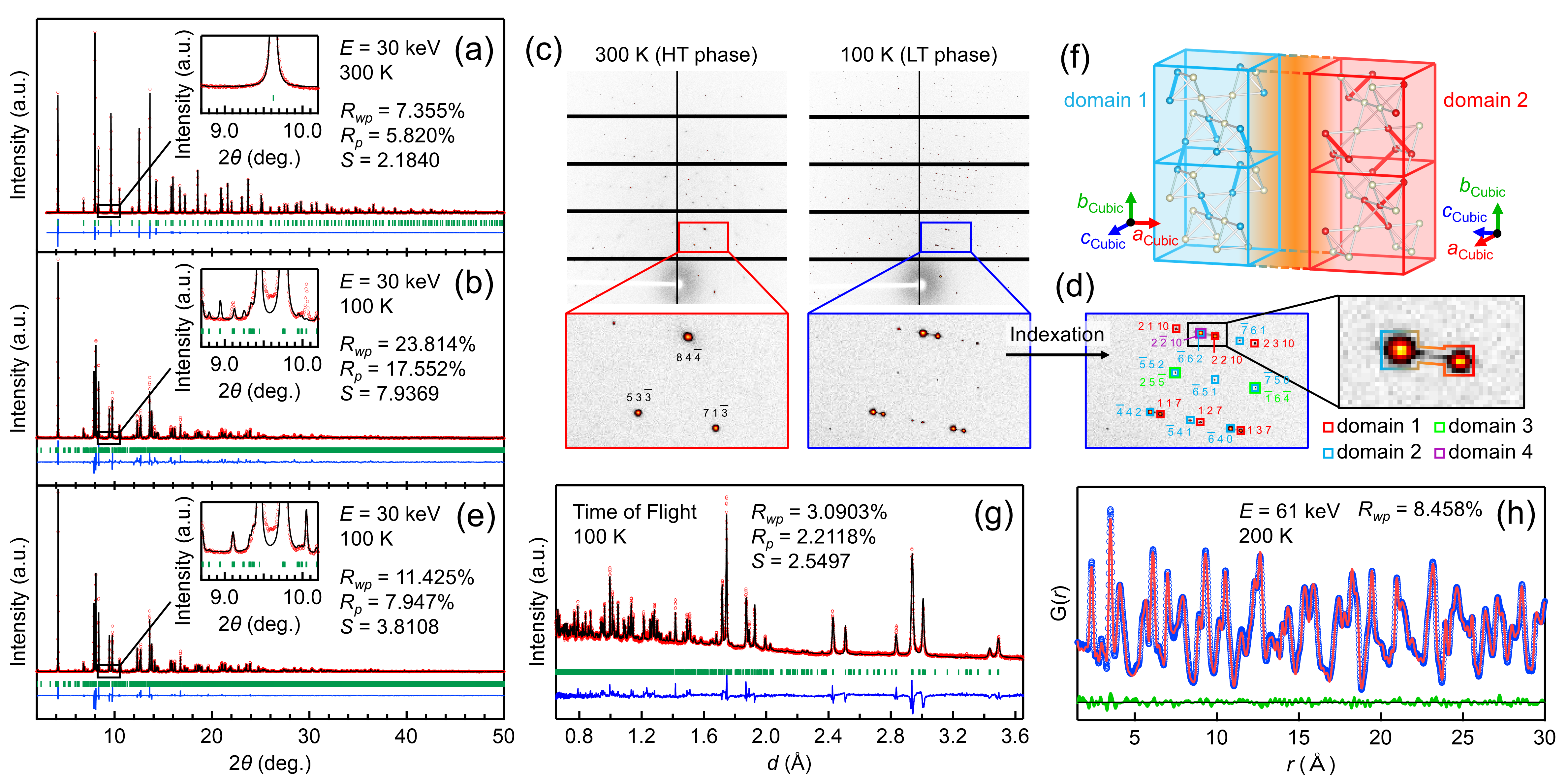}
\caption{\label{fig:Figure1} (a) Diffraction data at 300 K and the results of Rietveld analysis assuming cubic space group of $Fd\bar{3}m$. (b) Diffraction data at 100 K and the results of fitting using previously reported structural parameters \cite{CuIr2S4_radaelli}. (c) A part of single-crystal diffraction data at 300 K and 100 K. (d) Indexing of the diffraction data for the low-temperature phase assuming four domains. There is diffusion between the peaks that appear from different domains. (e) Results of Rietveld analysis using the structural parameters obtained from structural analysis by single-crystal x-ray diffraction at 100 K. (f) Relationship between the directions of the domains indicated by the indices in (d). In the actual system, there is no clear domain wall between these two domains, and the structure is thought to be gradually changing. (g) Results of Rietveld analysis using the structural parameters obtained from single-crystal structure analysis for the neutron diffraction data at 100 K. (h) Results of pair distribution function (PDF) analysis at 200 K using the structural parameters from the single-crystal structure analysis.}
\end{figure*}

In this paper, we report on the structural study of the low-temperature phase of CuIr$_2$S$_4$ using single-crystal and powder samples. Although multiple domains were observed in the low-temperature phase, we succeeded in clarifying the structure with a reliability factor of $R_1$ = 3.60\% by extracting the single domain component. A total of 253 parameters, including atomic coordinate parameters, anisotropic atomic displacement factors, and scale factors, were refined simultaneously, and they converged well. In the obtained structure, the lattice constants were almost the same as those reporeted previously \cite{CuIr2S4_radaelli}, but the arrangement of Ir$^{3+}$ and Ir$^{4+}$ on the $a$, $b$, and $c$ axes was different, and as a result, the arrangement of the dimers was also different. Similar crystal structures were independently obtained from the single-crystal structure analyses performed with different domain components. Rietveld analysis of synchrotron radiation powder x-ray diffraction data and neutron diffraction data was performed using the obtained structure as the initial structure, resulting in almost no change in the structural parameters and good $R$ values. Our results provide a new basis for studying CuIr$_2$S$_4$, which continues to offer many fascinating research themes in condensed matter physics, such as the effects of x-ray irradiation, novel magnetic properties that appear at low temperatures, short-range tetragonal distortion that appears at high temperatures, and superconductivity under high pressure.

\begin{table}[htb]
\caption{\label{tab:table1}Summary of structural analysis of CuIr$_2$S$_4$. Note that the theoretical all-electron energy per unit cell (Ry) obtained by structural optimization calculation was -623276.4012.}
  \begin{tabular}{|c|c|c|c|}  \hline
Temperature (K) & 100 (LT phase) & \multicolumn{2}{|c|}{300 (HT phase)}  \\ \hline
Wavelength (\AA) & 0.30940 & \multicolumn{2}{|c|}{0.30940} \\ \hline
Dimension ($\mu$m$^3$) & 60$\times$40$\times$30 & \multicolumn{2}{|c|}{60$\times$40$\times$30} \\ \hline
space group & $P\bar{1}$ & \multicolumn{2}{|c|}{$Fd\bar{3}m$} \\ \hline
$a$ (\AA) & 11.96453(30) & \multicolumn{2}{|c|}{9.85153(2)} \\ \hline
$b$ (\AA) & 6.99013(15) & \multicolumn{2}{|c|}{9.85153} \\ \hline
$c$ (\AA) & 11.94923(32) & \multicolumn{2}{|c|}{9.85153} \\ \hline
$\alpha$ ($^{\circ}$) & 91.0569(13) & \multicolumn{2}{|c|}{90} \\ \hline
$\beta$ ($^{\circ}$) & 108.4763(18) & \multicolumn{2}{|c|}{90} \\ \hline
$\gamma$ ($^{\circ}$) & 91.0323(13) & \multicolumn{2}{|c|}{90} \\ \hline
$V$ (\AA$^3$) & 947.3843(414) & \multicolumn{2}{|c|}{956.1163(28)} \\ \hline
$Z$ & 8 & \multicolumn{2}{|c|}{8} \\ \hline
$F$(000) & 1976 & \multicolumn{2}{|c|}{1976} \\ \hline
$d_{Min}$ (\AA) & 0.28 & \multicolumn{2}{|c|}{0.28} \\ \hline
$N_{Total, obs}$ & 168662 & \multicolumn{2}{|c|}{37197} \\ \hline
$N_{Unique, obs}$ & 66920 & \multicolumn{2}{|c|}{1207} \\ \hline
Average redundancy & 2.5 & \multicolumn{2}{|c|}{31.5} \\ \hline
Completeness & 0.728 & \multicolumn{2}{|c|}{0.993} \\ \hline
$R_1$ & 0.0360 & \multicolumn{2}{|c|}{0.0229} \\ \hline
$R_1$(superlattice peaks) & 0.0369 & \multicolumn{2}{|c|}{-} \\ \hline
$R_1$(fundamental peaks) & 0.0294 & \multicolumn{2}{|c|}{-} \\ \hline
GOF & 1.046 & \multicolumn{2}{|c|}{1.56} \\ \hline\hline
\multicolumn{4}{|c|}{Structural energy of LT phase strucure} \\ \hline
Report & Previous \cite{CuIr2S4_radaelli} & \multicolumn{2}{|c|}{This work} \\ \hline
Energy (Ry) & -623276.2187 & \multicolumn{2}{|c|}{-623276.3956} \\ \hline
  \end{tabular}
\end{table}

Powder samples of CuIr$_2$S$_4$ were synthesized by a solid-state reaction method: Cu, Ir, and S were mixed in stoichiometric ratios, vacuum-sealed in quartz tubes, and sintered at 850 $\tccentigrade$ for 7 days. The single-crystal samples were synthesized by the flux method using Bi-flux \cite{CuIr2S4_singlecrystal}. Powder x-ray diffraction measurements for Rietveld analysis were performed at BL02B2 and BL10XU \cite{BL10XU} at SPring-8 with an energy of $E$ = 30 keV. RIETAN-FP \cite{Rietan} was used for Rietveld analysis. Powder x-ray diffraction measurements for PDF analysis were performed at BL04B2 with an energy of $E$ = 61 keV. The reduced PDF $G$($r$) was obtained by the conventional Fourier transform \cite{PDF}. PDFgui package was used for analysis \cite{PDFgui}. Powder neutron diffraction measurements were performed at SuperHRPD at J-PARC \cite{SuperHRPD}, and Z-Rietveld \cite{ZRietveld1,ZRietveld2} was used for analysis. Single-crystal x-ray diffraction measurements were performed at BL02B1 at SPring-8 with an energy of $E$ = 40 keV. CrysAlisPro was used for indexing and intensity extraction\cite{crysalis}, SORTAV \cite{SORTAV} was used for outlier removal and data averaging, and SHELXL \cite{shelx}, which is built into WinGX \cite{WinGX}, and JANA2006 \cite{Jana} software used for structural refinement. The crystal structure was drawn using VESTA \cite{VESTA}. Magnetic susceptibility measurement was performed using the Quantum Design MPMS at the CROSS user laboratories. The calculations of energy and structure optimaization were performed using density-functional theory with the Perdew-Burke-Ernzerhof generalized-gradient approximation for solids (PBEsol) \cite{calc_1} and Kresse-Joubert projector-augmented-wave pseudopotentials \cite{calc_2,calc_3}, implemented in the \textsc{Quantum} ESPRESSO package \cite{calc_4,calc_5}.

The results of the powder diffraction experiment conducted at 300 K at the BL02B2 beamline of SPring-8 are shown in Figure~\ref{fig:Figure1}(a). Rietveld analysis was assuming a cubic spinel structure with the space group $Fd\bar{3}m$ showed good agreement with $R_{wp}$ = 7.383\%. When the temperature was lowered, the diffraction pattern clearly changed at about 230 K, as previously reported, confirming that a structural phase transition had occurred. As mentioned above, it has been reported previously that in this low-temperature phase, the space group is $P$$\bar{1}$ and that dimers are formed between adjacent Ir$^{4+}$ ions due to charge ordering of Ir. The results of fitting the diffraction data measured at 100 K to the structural parameters reported in a previous study are shown in Figure~\ref{fig:Figure1}(b) \cite{CuIr2S4_radaelli}. It should be noted that this temperature of 100 K is sufficiently high compared to 70 K, at which the x-ray irradiation effect appears \cite{CuIr2S4_xrayinduced}, and that no structural changes due to x-ray irradiation have occurred. As can be seen from the inset, in the current data, the position of the green vertical line indicating the peak coincides with the position of the peaks, including the superlattice reflections, but the intensity is clearly not reproduced. This indicates that the lattice constants are consistent with those reported previously, but the internal structure, such as atomic coordinations, differs from those reported previously \cite{CuIr2S4_radaelli}.

\begin{figure*}
\includegraphics[width=180mm]{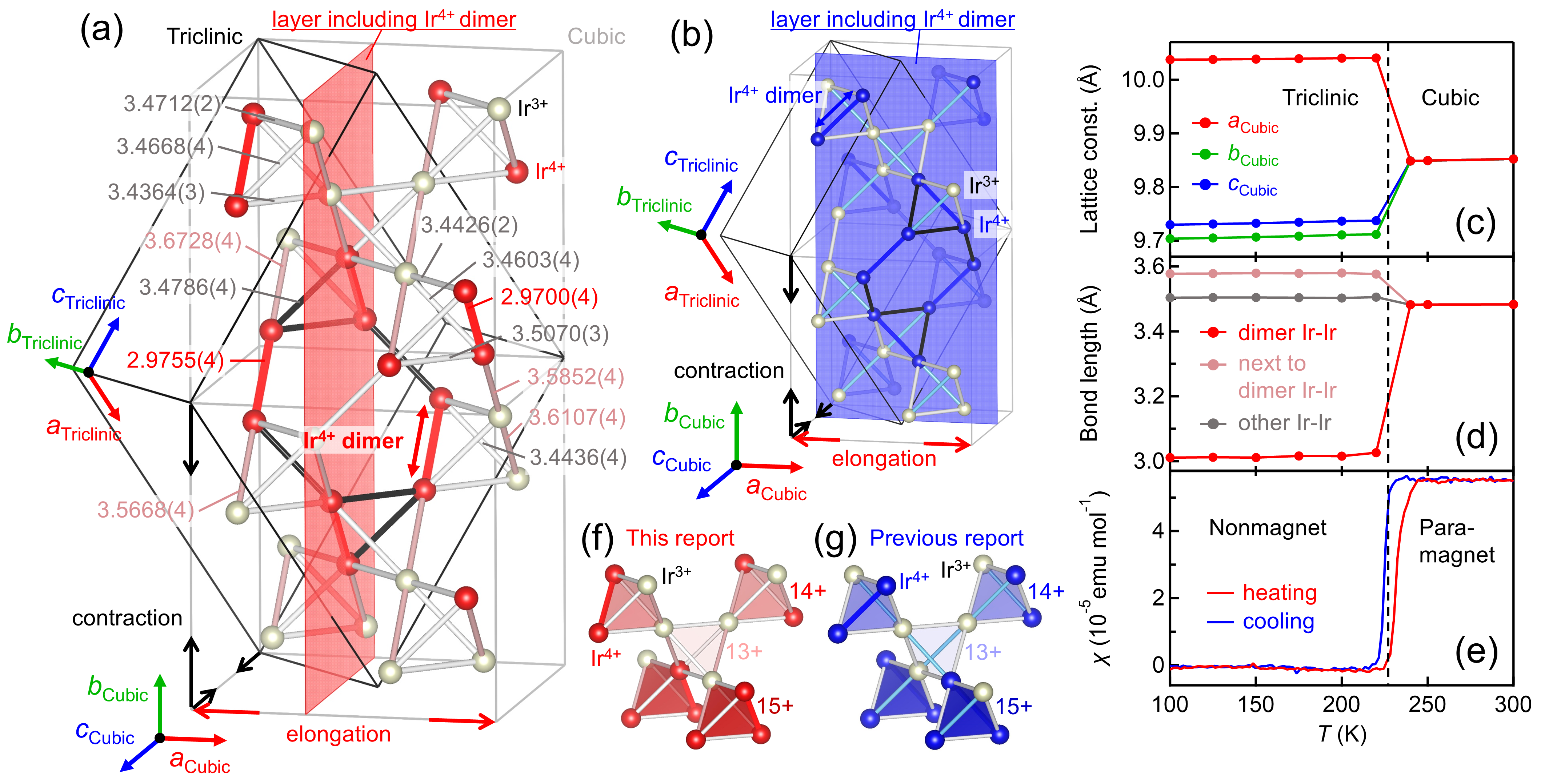}
\caption{\label{fig:Figure2} (a) Ir$^{3+}$/Ir$^{4+}$ charge ordering arrangement and Ir-Ir distances including dimer in the low-temperature phase of CuIr$_2$S$_4$ revealed in this experiment. (b) Charge ordering pattern and dimer arrangement reported in previous literature \cite{CuIr2S4_radaelli}. (c) Temperature dependence of the lattice constant. Since the direction of lattice distortion and the axis direction in the cubic setting are essentially the same, the lattice constants of the triclinic phase are converted to those of the cubic phase, $a$$_{\rm Cubic}$, $b$$_{\rm Cubic}$, $c$$_{\rm Cubic}$ using the following formula: $a$$_{\rm Cubic}$ = 1/2 $a$$_{\rm Triclinic}$ - $b$$_{\rm Triclinic}$ + 1/2 $c$$_{\rm Triclinic}$, $b$$_{\rm Cubic}$ = -1/2 $a$$_{\rm Triclinic}$ + 1/2 $c$$_{\rm Triclinic}$, $c$$_{\rm Cubic}$ = -1/2 $a$$_{\rm Triclinic}$ - $b$$_{\rm Triclinic}$ - 1/2 $c$$_{\rm Triclinic}$. The temperature dependence of lattice constants in the triclinic setting is shown in Appendix A. (d) Temperature dependence of Ir-Ir distance. There are 8 Ir sites in the unit cell, and there are 24 possible Ir-Ir distances. This graph shows the average of these Ir-Ir distances, divided into three groups. The color of each data point corresponds to the color of the bond shown in (a). (e) Temperature dependence of the magnetic susceptibility.  (f,g) Comparison of the total valence in the Ir tetrahedron of CuIr$_2$S$_4$ between this experiment and the previous report inside Ir tetrahedron.}
\end{figure*}

In order to determine the internal structure of the low-temperature phase, we conducted synchrotron x-ray diffraction experiments using single-crystal samples at the BL02B1 beamline of SPring-8. As shown in Figure~\ref{fig:Figure1}(c), in the 300 K, the diffraction peaks have a spotty shape, and there are no unknown peaks derived from impurities. When the temperature is lowered, a structural phase transition occurs at 230 K, and the number of peaks increases drastically, as can be seen on the right side of Figure~\ref{fig:Figure1}(c). A portion of the diffraction pattern obtained at 100 K is shown in Figure~\ref{fig:Figure1}(d). The diffraction data contains peaks derived from multiple domains that have appeared due to the low-symmetrization from $Fd\bar{3}m$ to $P\bar{1}$. At low temperatures, it is expected that up to 24 domains will form, but in this analysis, we were able to identify most of the high-intensity peaks by assuming only four domains. Some of the peaks in the low-temperature phase overlap with peaks from different domains, so it is difficult to accurately separate their peaks. On the other hand, there are also peaks that do not overlap with peaks from different domains, and their intensities can be accurately extracted. 

Therefore, we selected the domain with the most peaks among the four indexed and used only independent peaks that did not overlap with peaks from other one for structural analysis. Despite the reduction in the number of reflections available for analysis due to this procedure, it should be noted that this is still sufficient to refine the 253 parameters (atomic coordinates, anisotropic displacement factors, and scale factors) in the $P\bar{1}$ space group. After refining only the scale factor while leaving the previously reported structural parameters unchanged \cite{CuIr2S4_radaelli}, the $R_1$ value was 63.18\%, which is a poor value. Conversely, upon refining all parameters, including atomic positions and atomic displacement factors, the parameters demonstrated appropriate convergence, yielding an $R_1$ value of 3.60\%, which is considered a satisfactory outcome. The $R_1$ value calculated from the superlattice reflection alone is 3.69\%, and the $R_1$ value calculated from the main reflection alone is 2.94\% using JANA2006 \cite{Jana} software for refinement. Both of these are good values, indicating that the analysis has been carried out successfully. The results of the structural analysis are shown in Table~\ref{tab:table1}, and the structural parameters obtained are summarized in Tables in the Appendix B of this article. This analysis results were reproduced even when different initial structures were given or when intensity data of different domains were used. 

In the structural analysis using single-crystal, some peaks that overlap with peaks from other domains were not used. There may be inaccurate values in the structural parameters, so the structural parameters must be checked not only single-crystal x-ray diffraction, but also other structural analysis method. Using the structural parameters obtained from the single-crystal synchrotron x-ray diffraction experiment, the entire synchrotron powder x-ray diffraction data, including the superlattice peaks, were successfully fitted, as shown in Figure~\ref{fig:Figure1}(e) and its inset. Although there is an intensity that cannot be reproduced by the simulation around 2$\theta$ = 9.6$\tcdegree$, this does not mean that the assumed structural model is incorrect. This is due to the appearance of diffuse scatterings, as shown in Figure~\ref{fig:Figure1}(d). These diffuse scattering appears in various positions in the single-crystal diffraction data, and a common feature is that it connects two peaks that originate from different domains. This suggests that there is no clear domain wall between adjacent domains through a particular crystal plane indicated by the indices of the two peaks, and that the structures are gradually changing and connecting with finite thickness. For example, the index in Figure~\ref{fig:Figure1}(d) predicts that the relationship between the domains shown in the conceptual figure in Figure~\ref{fig:Figure1}(f) is being realized. Rietveld analysis cannot reproduce such diffuse scattering, so the misalignment shown in Figure~\ref{fig:Figure1}(e) occurs. Therefore, in the Rietveld analysis of the low-temperature phase, the $R$ value will inevitably be poor, so it is necessary to directly check the fitting for all peaks. In the analysis of Figure~\ref{fig:Figure1}(e), the intensities of many peaks, including the weak superlattice peaks, are well reproduced, indicating that the results of the structural analysis are correct.

The sulfur in the ligand has fewer electrons than iridium, so it makes a smaller contribution to the x-ray diffraction. Therefore, we performed a neutron diffraction experiment at 100 K to more accurately evaluate the structural parameters of the ligand. The results of fitting the neutron diffraction data with the structural parameters obtained from single-crystal x-ray diffraction are shown in Figure~\ref{fig:Figure1}(g). After refining the atomic positions, no significant changes from the given structure were observed. This indicates that the structural parameters determined in the single-crystal X-ray diffraction experiment were appropriate, including for sulfur. In addition, the fitting assuming the previously reported structural model \cite{CuIr2S4_radaelli} significantly reduced the accuracy of the analysis. In addition, as shown in Figure~\ref{fig:Figure1}(h), we confirmed that good fitting can be obtained for the PDF data when the crystal structure obtained from the single-crystal x-ray diffraction experiment is used as the initial structure.

The structure of the low-temperature phase obtained from this analysis is shown in Figure~\ref{fig:Figure2}(a). This arrangement of charge ordering pattern and dimer arrangement is clearly different from the arrangement shown in the previous report shown in Figure~\ref{fig:Figure2}(b). In previous reports, a structure was proposed in which a fourfold superstructure occurred in the Ir lattice that expanded in the $ab$ plane direction in the high-temperature phase cubic lattice, but in the current structure, a fourfold superstructure occurs in the Ir lattice that expands in the $bc$ plane direction. By calculating the theoretical total-electron energy of two structures shown in Figures~\ref{fig:Figure2}(a) and (b), it was found that the structure proposed in this study had lower energy and was more stable, as shown in Table~\ref{tab:table1}. Furthermore, when we performed a structural optimization calculation using Figure~\ref{fig:Figure2}(a) and (b) as initial structure, a structure similar to Figure~\ref{fig:Figure2}(a) became the energetically most stable structure, regardless of the initial structure. In fact, the energy obtained through structural optimization is very close to the energy of the structure that shown in Figure~\ref{fig:Figure2}(a). 

The reason for the large difference in energy between the two structures in Figure~\ref{fig:Figure2}(a) and (b) can be understood naturally when considering the relationship between the change in unit cell shape accompanying the phase transition and the arrangement of the dimers. When we examine the change in lattice constants accompanying the phase transition, using the $a$, $b$, and $c$ axes of the cubic crystal structure of the high-temperature phase as a reference, we can see that the $a$ axis has expanded greatly, while the $b$ and $c$ axes have contracted, as shown in Figure~\ref{fig:Figure2}(c). Since the interatomic distance shortens when dimers are formed, it is expected that the length of the axes in the direction in which the dimers form will shorten accompanying the phase transition. Therefore, it is predicted that the dimer is contained within the $bc$ plane of the high-temperature cubic crystal cell. In Figure~\ref{fig:Figure2}(a), the dimers are contained in the $bc$ plane, which is consistent with the change in lattice constant, while in Figure~\ref{fig:Figure2}(b), the dimers are contained in the $ab$ plane, which is inconsistent with the change in lattice constant. The above, together with the fact that the structure obtained in this study reproduces the diffraction pattern well, strongly supports the charge ordering and dimer arrangement obtained in this study as the ground state of CuIr$_2$S$_4$.

The changes in lattice constants and Ir-Ir distances associated with the sudden decrease in magnetic susceptibility are shown in Figure~\ref{fig:Figure2}(c)-(e). The structural change appears as a clear first-order phase transition, and as shown in Figure~\ref{fig:Figure2}(a), the Ir-Ir distances, which were all equal in the high-temperature phase, split into several different distances in the low-temperature phase. The Ir-Ir distances that form dimers show a contraction of about 15\% compared to the Ir-Ir distances in the high-temperature phase. This indicates that, in the low-temperature phase of CuIr$_2$S$_4$, there is a large gap in the electronic system that exceeds the energy loss due to the large lattice distortion.

While the charge ordering pattern and dimer arrangement proposed in this study are clearly different from those reported previously \cite{CuIr2S4_radaelli}, it should be noted that the important features of charge ordering on the pyrochlore lattice are still present in the present structure. Figures~\ref{fig:Figure2}(f) and (g) show the results of calculating the total sum of the Ir valence within each Ir$_4$ tetrahedron that constitutes the pyrochlore lattice. Figure~\ref{fig:Figure2}(f) shows the results of calculations based on the structure proposed in Figure~\ref{fig:Figure2}(a), and Figure~\ref{fig:Figure2}(g) shows the results of calculations based on the structure reported previously \cite{CuIr2S4_radaelli} in Figure~\ref{fig:Figure2}(b). In both Figure~\ref{fig:Figure2}(f) and (g), the total charge in the tetrahedron varies between 13+ and 15+, indicating that the Anderson condition for charge ordering is not satisfied \cite{anderson, CuIr2S4_radaelli}. As a reason for the realization of a charge ordered state that does not satisfy the Anderson condition, it has been proposed that in the structure shown in Figure~\ref{fig:Figure2}(b), which was reported earlier, the increase in energy due to the formation of the $\sigma$ bond between Ir$^{4+}$ is so large that the loss of Coulomb energy due to the violation of the Anderson condition is negligible \cite{anderson_discussion,anderson_discussion2}. This basic idea should also be applicable to the present structure. On the other hand, it has been reported that in LiRh$_2$O$_4$, which has the same $d^{5.5}$ electronic state as CuIr$_2$S$_4$, a charge ordered state and a dimer state that satisfy the Anderson condition appear at low temperatures \cite{LiRh2O4}. There must be some reason why the Anderson condition is not satisfied in CuIr$_2$S$_4$, which is similar to LiRh$_2$O$_4$ in its ground state, such as the influence of spin-orbit interaction due to 5$d$ electrons.

 \begin{figure}
\includegraphics[width=85mm]{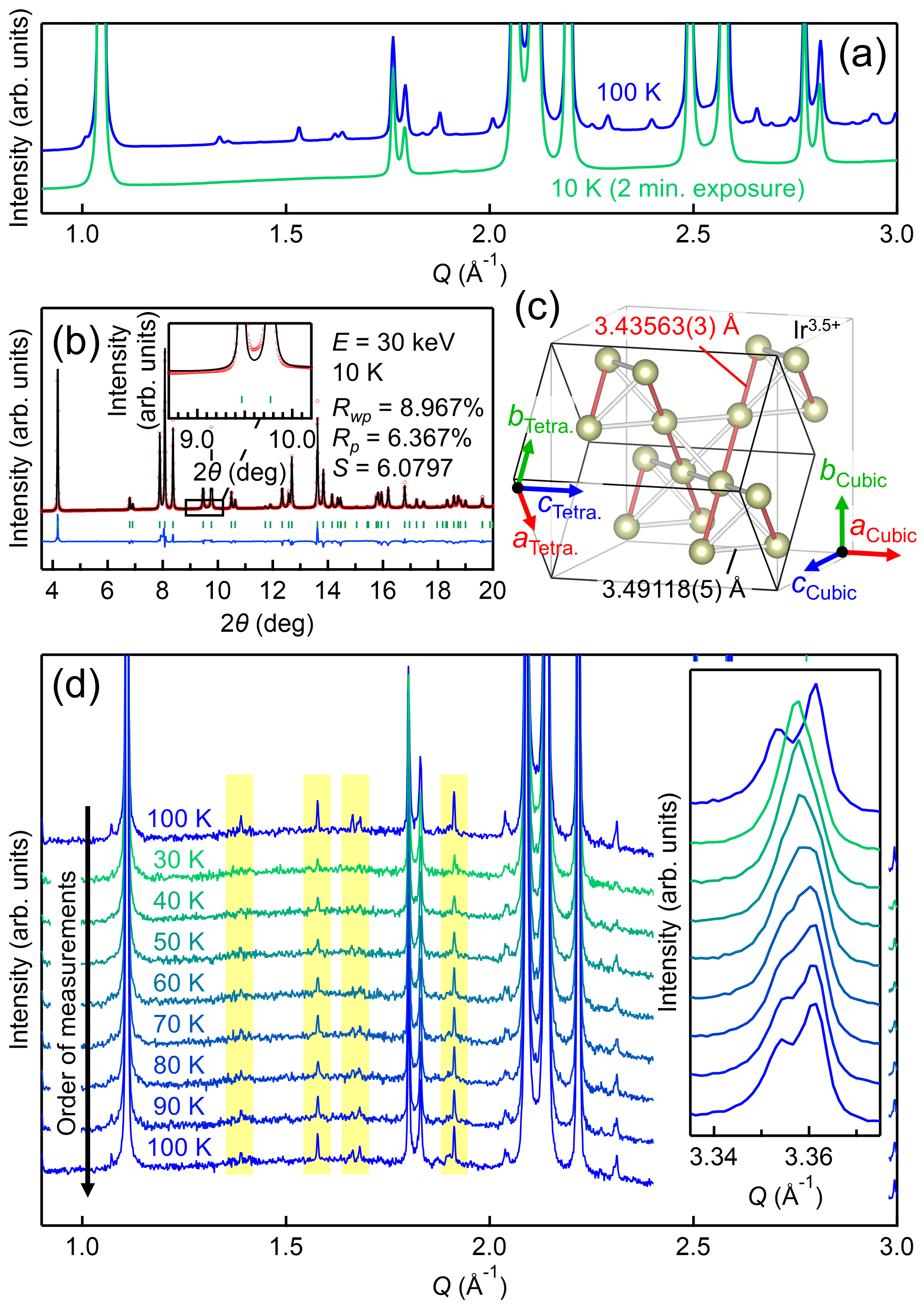}
\caption{\label{fig:Figure3} (a) A portion of the synchrotron x-ray powder diffraction patterns measured at 100 K and 10 K at the SPring-8 BL10XU beamline. (b) The results of Rietveld analysis assuming the tetragonal space group $I$4$_1$/$amd$ for the data measured at 10 K. (c) The tetragonal structure with the disappearance of the charge order and dimerization achieved after high-intensity x-ray irradiation at 10 K. (d) Diffraction patterns on the slowly heating process at SPring-8 BL02B2 beamline. Inset shows change of the fundamental peaks. Diffraction patterns at 30 K were measured after irradiating x-ray for 86 minutes.}
\end{figure}


Finally, as an example of the importance of the clarification of the low-temperature phase structure, we would like to introduce x-ray irradiation effect that appears at temperatures below 70 K. Figure~\ref{fig:Figure3}(a) shows the results of a 10 K diffraction experiment on the BL10XU beamline at SPring-8. The experiment used detuned undulator radiation with an energy of $E$ = 30 keV. In contrast to the results of a measurement conducted at 100 K using the same energy of light at the BL02B2 beamline at SPring-8, the superlattice reflections completely disappeared and some of the fundamental reflections merged. This indicates that the high-intensity x-ray irradiation has melted the charge-ordered dimer molecules and increased the symmetry, as mentioned in the introduction. As shown in Figure~\ref{fig:Figure3}(b) and (c), the diffraction pattern after the disappearance of the superlattice can be refined by Rietveld analysis assuming the tetragonal space group of $I$4$_1$/$amd$, as reported previously \cite{CuIr2S4_xrayinduced}. As shown in Figure~\ref{fig:Figure3}(d), when the temperature is increased, the superlattice peaks reappear and the fundamental reflections, which had been aggregated, show a tendency to split again. This indicates that the lattice is becoming less symmetrical, from tetragonal to triclinic. The reappearance of the triclinic phase, which had been suppressed at the lowest temperature, seems to indicate that the spontaneous generation of the charge order/dimer-containing superstructure, which had been kinetically suppressed, is accelerated by the increase in temperature. In fact, in past research where samples were fixed at various temperatures after being irradiated with x-rays at the lowest temperature, it was reported that if they were maintained at higher temperatures, the electrical resistivity would rapidly return to the value it had before irradiation \cite{CuIr2S4_xrayinduced2}. This behavior is reminiscent of the light-induced excited spin state trapping (LIESST) effect in spin transitions \cite{LIESST1,LIESST2}.

In this experiment, the superlattice peak completely disappeared after 120 seconds of exposure time, so it was not possible to investigate the melting process of the charge order or dimer. This is probably due to the intensity of the x-rays being too strong. In the previously reported paper \cite{CuIr2S4_xrayinduced}, the change in the diffraction pattern accompanying x-ray irradiation was captured as a time-dependent phenomenon, and if such an experiment could be performed, the low-temperature phase structure revealed in this study should play an important role. In other words, by performing Rietveld analysis on the data for the melting process accompanying irradiation, it should be possible to track the melting process of the charge order and dimers as a time-dependent phenomenon. Similarly, by investigating the changes in the structure under high pressure using the structural data proposed in this study as the initial structure, we should be able to obtain important clues towards elucidating the mechanism of the high-$T_c$ superconductivity phenomenon under high pressure recently reported in CuIr$_2$S$_4$. In this way, CuIr$_2$S$_4$ is an important material that continues to provide fascinating topics in the field of condensed matter physics, and the results of this research will undoubtedly provide an important foundation for studying it.

In summary, we have investigated the crystal structure of CuIr$_2$S$_4$ at low temperatures using single-crystal synchrotron x-ray diffraction experiments, and have clarified that a charge order and dimer arrangement different from those reported previously \cite{CuIr2S4_radaelli} are realized. The results of this study should provide an important basis for research on CuIr$_2$S$_4$, which continues to generate many new research interests.

\begin{acknowledgments}
The authors are grateful to Jun-ichi Yamaura for the experimental support and to Takashi Mizokawa for valuable discussion. This work was carried out under the Visiting Researcher’s Program of the Institute for Solid State Physics, the University of Tokyo, and the Collaborative Research Projects of Laboratory for Materials and Structures, Institute of Innovative Research, Tokyo Institute of Technology. The work leading to these results has received funding from the Grant in Aid for Scientific Research (Nos.~JP20H02604,  JP21J21236, JP21K18599, JP22KJ1521, JP23H04104, JP24H01620, JP24K01329, JP23K03286) and JST COI-NEXT Program (Nos.~JPMJPF2221). Synchrotron XRD experiment was performed at BL02B1, BL02B2, BL04B2 and BL10XU (Proposals No. 2022A1098, No. 2022B1130, No. 2023A1110, No. 2023A1869, No. 2023B0304, No. 2023B1114, No. 2023B1147, No. 2023B1921, No. 2024A0304, No. 2024A1151, No. 2024A1704, No. 2024B0304) equipped at SPring-8, Hyogo, Japan. Powder x-ray diffraction experiments were performed on BL5S2 at Aichi SR. (Proposals No. 202203061, No. 202301056, No. 202306067, No. 202403098). A neutron diffraction experiment was conducted at the SuperHRPD of J-PARC, Tokai, Japan (Proposal No. 2024B0015). 
\end{acknowledgments}

\begin{appendix}

\section{Temperature dependence of lattice constants (in triclinic cell)}
The conversion of Triclinic to Cubic, described in the caption of Figure~\ref{fig:Figure2}(c), is not strictly correct (the angles of $\alpha$$_{\rm Cubic}$, $\beta$$_{\rm Cubic}$, and $\gamma$$_{\rm Cubic}$ deviate slightly from 90 degrees). In this term, we converted Cubic to Triclinic. However, this conversion does not show the direction of lattice distortion at a glance, so the conversion of Triclinic to Cubic is used in Figure~\ref{fig:Figure2}(c).

\section{single-crystal x-ray diffraction analysis results}
The results of this structural analysis are shown in the following table. The details of the experimental conditions and the results obtained can also be obtained free of charge from the Cambridge Crystallographic Data Centre (https://www.ccdc.cam.ac.uk/) as a crystal information file (CIF). The supplementary crystal structure data for this paper is summarized in CSD 2434099.

 \begin{figure}
\includegraphics[width=85mm]{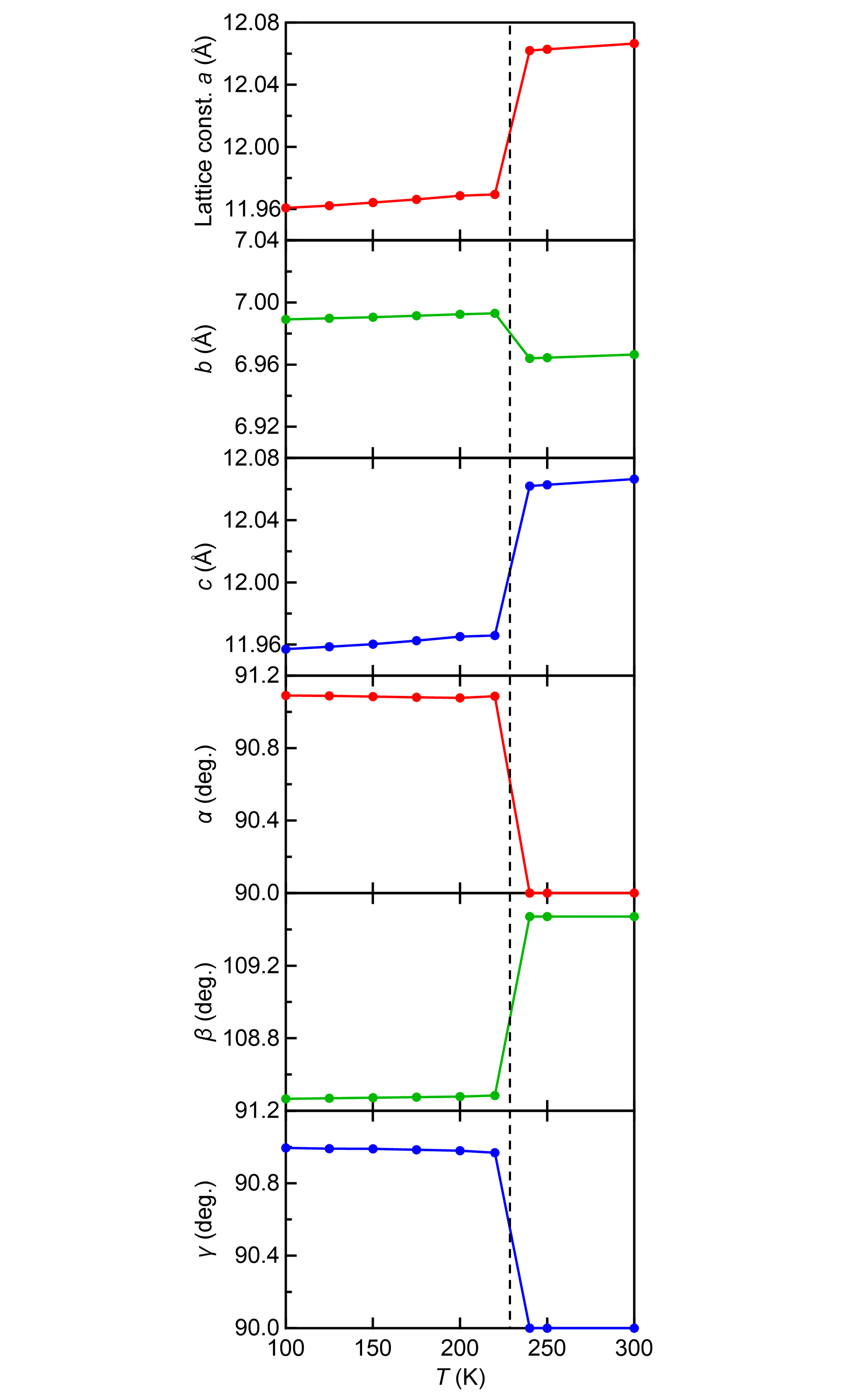}
\caption{\label{fig:Figure4} Temperature dependence of the lattice constants estimated in the triclinic setting using the following formula: $a$$_{\rm Triclinic}$ = 1/2 $a$$_{\rm Cubic}$ - $b$$_{\rm Cubic}$ - 1/2 $c$$_{\rm Cubic}$, $b$$_{\rm Triclinic}$ = -1/2 $a$$_{\rm Cubic}$ - 1/2 $c$$_{\rm Cubic}$, $c$$_{\rm Triclinic}$ = 1/2 $a$$_{\rm Cubic}$ + $b$$_{\rm Cubic}$ - 1/2 $c$$_{\rm Cubic}$.}
\end{figure}

\begin{table*}[htb]
\centering

\caption{\label{tab:table2}%
Structural parameters of {CuIr}$_2${S}$_4$ at 300 K. 
}
\begin{ruledtabular}
\begin{tabular}{cccccc}
&\multicolumn{5}{c}{atomic coordinates}\\
site & Wyck. & Occ. & $x/a$ & $y/b$ & $z/c$ \\ 
\hline
Cu&8$a$&1&1/8&1/8&1/8\\
Ir&16$d$&1&1/2&1/2&1/2\\
S&32$e$&1&0.73967(3)&0.73967&0.73967\\
\end{tabular}
\end{ruledtabular}
\leftline{ }
\leftline{ }

\end{table*}

\begin{table*}[htb]
\caption{\label{tab:table3}%
Anisotropic atomic displacement parameters of {CuIr}$_2${S}$_4$ at 300 K. 
}
\begin{ruledtabular}
\begin{tabular}{ccccccc}
site~~ & $U_{11}$ (\AA$^{2}$) & $U_{22}$ (\AA$^{2}$) & $U_{33}$ (\AA$^{2}$) & $U_{12}$ (\AA$^{2}$) & $U_{13}$ (\AA$^{2}$) & $U_{23}$ (\AA$^{2}$) \\ \hline
Cu&0.01064(4)&0.01064&0.01064&0&0&0\\
Ir&0.005651(16)&0.005651&0.005651&-0.000328(3)&-0.000328&-0.000328\\
S&0.00637(3)&0.00637&0.00637&-0.00006(3)&-0.00006&-0.00006\\

\end{tabular}
\end{ruledtabular}
\end{table*}

\begin{table*}[htb]
\caption{\label{tab:table4}%
Structural parameters of {CuIr}$_2${S}$_4$ at 100 K. The Ir atoms at Ir1-Ir4 sites are involved in dimer formation, while the Ir atoms at Ir5-Ir8 sites are isolated. Therefore, it is thought that the Ir atoms at Ir1-Ir4 sites are 4+ and the Ir atoms at Ir5-Ir8 sites are 3+.
}
\begin{ruledtabular}
\begin{tabular}{cccccc}
&\multicolumn{5}{c}{atomic coordinates}\\
site & Wyck. & Occ. & $x/a$ & $y/b$ & $z/c$ \\ 
\hline
Cu1&2$i$&1&0.06527(4)&0.25358(7)&0.93632(4)\\
Cu2&2$i$&1&0.43529(4)&0.27026(7)&0.06689(4)\\
Cu3&2$i$&1&0.94046(4)&0.26055(7)&0.55921(4)\\
Cu4&2$i$&1&0.56207(4)&0.27915(7)&0.44167(4)\\
Ir1&2$i$&1&0.73131(2)&0.99070(2)&0.25265(2)\\
Ir2&2$i$&1&0.25194(2)&0.01929(2)&0.26375(2)\\
Ir3&2$i$&1&0.24977(2)&0.22747(2)&0.48277(2)\\
Ir4&2$i$&1&0.48736(2)&0.21674(2)&0.74581(2)\\
Ir5&2$i$&1&0.99288(2)&0.24176(2)&0.24918(2)\\
Ir6&2$i$&1&0.74412(2)&0.23211(2)&0.99775(2)\\
Ir7&2$i$&1&0.74563(2)&0.49875(2)&0.25006(2)\\
Ir8&2$i$&1&0.25257(2)&0.51541(2)&0.25035(2)\\
S1&2$i$&1&0.86747(9)&0.46709(14)&0.12839(9)\\
S2&2$i$&1&0.85877(9)&0.01290(14)&0.13108(9)\\
S3&2$i$&1&0.37280(9)&0.99905(14)&0.13639(9)\\
S4&2$i$&1&0.37455(9)&0.54950(14)&0.13202(9)\\
S5&2$i$&1&0.88929(9)&0.99008(14)&0.63757(9)\\
S6&2$i$&1&0.87109(9)&0.52859(15)&0.63003(9)\\
S7&2$i$&1&0.36963(9)&0.45376(13)&0.62526(9)\\
S8&2$i$&1&0.36710(9)&0.99769(15)&0.60719(9)\\
S9&2$i$&1&0.36754(9)&0.26750(15)&0.86597(9)\\
S10&2$i$&1&0.14784(9)&0.24567(15)&0.62621(9)\\
S11&2$i$&1&0.13093(9)&0.25427(14)&0.14327(9)\\
S12&2$i$&1&0.36603(8)&0.28028(14)&0.36211(9)\\
S13&2$i$&1&0.62663(9)&0.23440(14)&0.64058(9)\\
S14&2$i$&1&0.85688(8)&0.23608(14)&0.86610(9)\\
S15&2$i$&1&0.85970(9)&0.24807(14)&0.35685(9)\\
S16&2$i$&1&0.63478(9)&0.24264(15)&0.13121(9)\\

\end{tabular}
\end{ruledtabular}
\end{table*}

\begin{table*}[htb]
\caption{\label{tab:table5}%
Anisotropic atomic displacement parameters of {CuIr}$_2${S}$_4$ at 100 K. 
}
\begin{ruledtabular}
\begin{tabular}{ccccccc}
site~~ & $U_{11}$ (\AA$^{2}$) & $U_{22}$ (\AA$^{2}$) & $U_{33}$ (\AA$^{2}$) & $U_{12}$ (\AA$^{2}$) & $U_{13}$ (\AA$^{2}$) & $U_{23}$ (\AA$^{2}$) \\ \hline
Cu1&0.00546(9)&0.00463(9)&0.00529(9)&0.00011(7)&0.00204(8)&0.00018(7)\\
Cu2&0.00451(8)&0.00431(9)&0.00496(8)&0.00034(7)&0.00173(7)&0.00022(6)\\
Cu3&0.00538(9)&0.00454(9)&0.00476(9)&0.00015(7)&0.00188(7)&0.00011(7)\\
Cu4&0.00436(8)&0.00432(9)&0.00468(8)&0.00019(6)&0.00159(7)&0.00012(6)\\
Ir1&0.00306(2)&0.00265(2)&0.00316(2)&0.00015(2)&0.00127(2)&0.00012(1)\\
Ir2&0.00292(2)&0.00258(2)&0.00319(2)&0.00013(1)&0.00121(2)&0.00026(1)\\
Ir3&0.00294(2)&0.00272(2)&0.00307(2)&0.00011(2)&0.00113(2)&0.00014(1)\\
Ir4&0.00288(2)&0.00278(2)&0.00315(2)&0.00009(2)&0.00117(2)&0.00026(2)\\
Ir5&0.00274(2)&0.00274(2)&0.00318(2)&0.00011(2)&0.00114(2)&0.00016(2)\\
Ir6&0.00295(2)&0.00270(2)&0.00293(2)&0.00002(2)&0.00112(2)&0.00016(2)\\
Ir7&0.00300(2)&0.00252(2)&0.00315(2)&0.00001(2)&0.00127(2)&0.00025(2)\\
Ir8&0.00292(2)&0.00256(2)&0.00310(2)&0.00012(2)&0.00122(2)&0.00009(1)\\
S1&0.00391(19)&0.00361(19)&0.00420(20)&0.00014(15)&0.00164(16)&0.00002(14)\\
S2&0.00410(19)&0.00370(20)&0.00430(20)&0.00021(15)&0.00159(16)&0.00037(15)\\
S3&0.00386(19)&0.00360(20)&0.00402(19)&0.00022(15)&0.00149(16)&0.00020(14)\\
S4&0.00389(19)&0.00370(20)&0.00372(19)&0.00006(14)&0.00131(16)&0.00023(14)\\
S5&0.00391(19)&0.00365(19)&0.00427(19)&0.00021(15)&0.00155(16)&0.00021(15)\\
S6&0.00410(20)&0.00360(20)&0.00390(20)&0.00012(15)&0.00145(16)&0.00029(15)\\
S7&0.00386(19)&0.00345(19)&0.00410(20)&0.00000(15)&0.00152(16)&0.00003(15)\\
S8&0.00430(20)&0.00350(20)&0.00410(20)&-0.00014(15)&0.00156(16)&0.00018(15)\\
S9&0.00414(19)&0.00370(20)&0.00410(19)&0.00000(15)&0.00172(16)&0.00022(15)\\
S10&0.00399(19)&0.00360(20)&0.00440(20)&0.00011(15)&0.00146(17)&0.00017(15)\\
S11&0.00378(19)&0.00400(20)&0.00420(20)&0.00016(15)&0.00149(16)&0.00027(15)\\
S12&0.00359(18)&0.00366(19)&0.00406(19)&0.00015(15)&0.00120(16)&-0.00004(14)\\
S13&0.00410(20)&0.00360(20)&0.00400(20)&0.00007(15)&0.00155(17)&0.00023(15)\\
S14&0.00371(19)&0.00380(20)&0.00419(19)&0.00022(15)&0.00146(16)&0.00015(15)\\
S15&0.00385(19)&0.00360(20)&0.00400(19)&0.00012(15)&0.00141(16)&0.00013(15)\\
S16&0.00400(20)&0.00340(20)&0.00403(19)&0.00018(15)&0.00150(16)&0.00022(15)\\

\end{tabular}
\end{ruledtabular}
\end{table*}



\begin{table*}[htb]
\caption{\label{tab:table6}%
The space group and lattice constants by structure optimization. 
}
\begin{ruledtabular}
\begin{tabular}{cccccccc}
Space group & $a$ (\AA) & $b$ (\AA) & $c$ (\AA) & $\alpha$ ($^{\circ}$) & $\beta$ ($^{\circ}$) & $\gamma$ ($^{\circ}$) & $V$ (${\AA}^3$) \\ 
\hline
$P\bar{1}$&11.935&6.959&11.915&90.868&108.639&90.8228&937.35\\

\end{tabular}
\end{ruledtabular}
\end{table*}

\begin{table*}[htb]
\caption{\label{tab:table7}%
Structural parameters of {CuIr}$_2${S}$_4$ by structure optimization. The structure obtained is similar to the results of the structural analysis in this study, and the Ir atoms at the Ir1-Ir4 sites are involved in the formation of dimers, while the Ir atoms at the Ir5-Ir8 sites are isolated.
}
\begin{ruledtabular}
\begin{tabular}{cccccc}
&\multicolumn{5}{c}{atomic coordinates}\\
site & Wyck. & Occ. & $x/a$ & $y/b$ & $z/c$ \\ 
\hline
Cu1&2$i$&1&0.06514&0.25205&0.93692\\
Cu2&2$i$&1&0.43531&0.26728&0.06628\\
Cu3&2$i$&1&0.94079&0.25802&0.56003\\
Cu4&2$i$&1&0.56196&0.27571&0.44110\\
Ir1&2$i$&1&0.73331&0.99197&0.25255\\
Ir2&2$i$&1&0.25233&0.01549&0.26218\\
Ir3&2$i$&1&0.25053&0.22862&0.48559\\
Ir4&2$i$&1&0.48902&0.22009&0.74646\\
Ir5&2$i$&1&0.99400&0.24217&0.24969\\
Ir6&2$i$&1&0.74446&0.23433&0.99821\\
Ir7&2$i$&1&0.74583&0.49861&0.24932\\
Ir8&2$i$&1&0.25236&0.51332&0.25061\\
S1&2$i$&1&0.86845&0.47006&0.12818\\
S2&2$i$&1&0.86073&0.01325&0.13091\\
S3&2$i$&1&0.37225&0.99698&0.13491\\
S4&2$i$&1&0.37407&0.54473&0.13213\\
S5&2$i$&1&0.88745&0.99077&0.63620\\
S6&2$i$&1&0.87154&0.52634&0.63000\\
S7&2$i$&1&0.36913&0.45817&0.62594\\
S8&2$i$&1&0.36806&0.99853&0.60941\\
S9&2$i$&1&0.36816&0.26576&0.86708\\
S10&2$i$&1&0.14636&0.24804&0.62672\\
S11&2$i$&1&0.13054&0.25381&0.14224\\
S12&2$i$&1&0.36627&0.27626&0.36329\\
S13&2$i$&1&0.62692&0.23797&0.64009\\
S14&2$i$&1&0.85822&0.23810&0.86744\\
S15&2$i$&1&0.86165&0.24788&0.35800\\
S16&2$i$&1&0.63463&0.24340&0.13078\\

\end{tabular}
\end{ruledtabular}
\end{table*}




\end{appendix}

\clearpage

\nocite{*}

\bibliography{references}

\end{document}